\documentclass{article}

\usepackage[cp1251]{inputenc}

\usepackage[english,russian]{babel}

\usepackage{amssymb}
\usepackage{amsmath}

\usepackage[dvips]{graphicx}
\begin{document}

%%% remove comment delimiter ('%') and select language if required
\selectlanguage{english} 

\title{On some possible features of motion of a polaron in gyrotropic medium}

\author{Alexander Severin \thanks{The author would like to thank Igor Todoshchenko for the discussion and assistance in preparing the English version of the paper.} \\ \texttt{severin@kiam.ru}}

\begin{abstract}
We predict the possibility of asymmetric dynamics of polaron in gyrotropic medium and give approximate quantitative estimate of the effect.
\end{abstract}

\maketitle

There is a class of phenomena of charge transport in solids, which is theoretically possible, but so far hardly has been studied experimentally. This is a spatially asymmetric electrical conductivity. The effect of Bloch electrons on the lattice is weak so that their motion can be regarded as motion in a potential field, and therefore the band conductivity in opposite directions is the same. However, this approach is not applicable for ion and polaron conductivities. Ions and polarons cause significant deformation of the lattice due to their large mass. Therefore, ion and polaron conductivities may be different in opposite directions.

In the whole history of experiments on the subject the diode properties of a homogeneous medium have been observed, to our knowledge, only by Gurevich and Zheludev \cite{1} and Kanaev and Malinovsky \cite{2}. However, Kanaev and Malinovsky explained the effect by the presence of microdomains of ferroelectric, i.e. by violation of homogeneity. Gurevich and Zheludev have found that the conductivity of lithium sulphate monohydrate along one of the axes is asymmetric with respect to the reversal of direction. The otherwise linear I-V curve had about 10\% kink at the origin. The charge carriers were likely hydrogen ions moving by Grotthuss (the relay) mechanism.

Nevertheless, asymmetric dynamics of polarons has not been experimentally detected anywhere.  Let us realize what should be the substance is which this phenomenon could be detected. First of all, the crystal structure of this material should be gyrotropic, i. e., there must be at least one plane, such that the structure reflected with respect to this plane could not be translated to the original one by any rotation or movement. These are the same requirements for symmetry as that of the pyroelectric, and they correspond to the following ten point groups that do not have a center of symmetry: 1, 2, 3, 4, 6, m, mm2, 3m, 4mm, 6mm. Most likely, it will be a molecular crystal because in such crystals the charge transport is often due to small polarons.

The asymmetry of the polaron interaction with the media can manifest itself in many ways, but the easiest asymmetry to quantify is the one caused by the quadratic nonlinear dielectric polarization.

In a nonlinear dielectric medium polarization $p$ is given by:

\[p=\chi E+\chi ^{(2)} E^{2} +\chi ^{(3)} E^{3} +...\]

\noindent where $E$ - electric field, $\chi$, $\chi^{(2)}$, $\chi^{(3)}$ - linear, quadratic and cubic polarizabilities. Materials with quadratic and cubic polarization are widely used in optics for the second harmonic generation. Quadratic polarization is also present in all pyroelectrics.

It is easy to see that if the charge is located in a linear dielectric medium, the interaction  with the charge dipoles environment is fully balanced. However, if the environment is non-linear quadratic -- it is not.

The strength of the interaction of a polaron with a quadratic nonlinear medium is:

\[F=\int _{R}\chi ^{(2)} E_{x}^{2} \frac{\partial E_{x} }{\partial x} dr \]

\noindent where $R$ - the region of space outside the polaron, $x$ - the direction in which the medium quadratic nonlinear.

If we assume that the linear component of the polarizability is isotropic, and the polaron has a spherical shape, then:

\begin{equation} \label{1} 
F=\frac{4\pi }{15} \left(\frac{e}{4\pi \varepsilon \varepsilon _{0} } \right)^{3} \frac{\chi ^{(2)} }{r_{0}^{4} } .  
\end{equation}

\noindent where $e$ - elementary charge, $\varepsilon$ - the dielectric constant of the medium, $\varepsilon_{0}$ - dielectric constant of vacuum, $r_{0}$ - radius of the polaron.

The action of force \eqref{1} on a single polaron is equivalent to the effect of external electric field $E^{*}$:

\begin{equation} \label{2} 
E^{*} =\frac{4\pi }{15} \frac{e^{2} }{\left(4\pi \varepsilon \varepsilon _{0} \right)^{3} } \frac{\chi ^{(2)} }{r_{0}^{4} } .  
\end{equation}

We should realize that the formula \eqref{1} is correct only for large enough polarons. The first and most obvious reason is that at distances of the order of the lattice constant the dielectric medium  can not be considered as a homogeneous. However, there is another, no less important one. Movement of the polaron to the next location is due to thermal fluctuations of the lattice which reduce or even eliminate the potential barrier between locations. These fluctuations should strongly influence the charge distribution near the polaron, but could they destroy also the asymmetry arisen as a result of the polarization of the medium? 

Let us denote the energy of the polaron in this location in the absence of lattice vibrations $W_{1}$ and energies in two adjacent locations in opposite directions $W_{0}$ and $W_{2}$. In the crystal with a center of symmetry $W_{0}=W_{2}$. However, in noncentrosymmetric crystal, because of the effect described above, $W_{0}\neq W_{2}$, namely $W_{0}-W_{2} = 2Fa$, where $a$ is the lattice period. 

Now we are interested in the relative probability of fluctuations kicking polaron from location 1 to locations 0 and 2. If this energy difference would be created by the external field, we would say that the jump to the location 2 is more likely because the energy barrier is lower. However, in our case the energy difference is created by oscillating charges of the lattice itself. We can thus guess that the asymmetry of polarization near the polaron disappears, but the field produced by a part of the lattice lying on some fairly large distance will still act as an external. 

Equilibrium states of the polaron in locations 0, 1 and 2 have the same energy due to the translational invariance. Polaron transition from location 1 to location 0 is a time-reversed transition from location 1 to location 2. If we neglect the phonon scattering, the interatomic interaction field can be considered as the potential one. The electric field of the electron whose localization resulted in the formation of polaron can also be considered as a potential field, since the mass of the electron is much smaller than the mass of the atom, and therefore the state of the electron is a function of the atomic coordinates. Thus, within this approximation the jump of the polaron is a motion in a potential field, which is reversible. Consequently, jumps in the location 0 and 2 require the same energy and are equally likely. 

Asymmetry is possible only beyond the described approach, that is, if we do not neglect the scattering of phonons and electron mass. The first of these factors is convenient to consider using the phonon mean free path $l_{ph}$ approach.

Phonons created due to the jump of the polaron to a new location do not go beyond the sphere of radius $l_{ph}$. In turn, phonons which kick polaron from old location must be born as a result of fluctuations within a radius $l_{ph}$. Thus, the entire process of polaron hopping, which we considered approximately reversible above, is localized a sphere with radius $l_{ph}$, and the media outside can be treated as the source of the external field. Therefore, we can use the formulas \eqref{1} and \eqref{2} as a rough estimate also for small polarons if we substitute phonon mean free path $l_{ph}$ instead of the polaron radius $r_{0}$. 

Let us estimate the effect of anisotropy on a particular substance, meta-nitroaniline (3-nitroaniline, $\mu$-nitroaniline) - C${}_{6}$H${}_{6}$N${}_{2}$O${}_{2}$. Meta-nitroaniline crystallizes in the form of molecular crystal with orthorhombic lattice and four molecules in the elementary cell, and the unit cell dimensions are $a$=6.501 \AA, $b$=19.33 \AA, $c$=5.082 \AA. \cite{5}

Monocrystalline meta-nitroanilin has a high quadratic nonlinearity. The mechanism of conductivity is hopping, which is proved by its temperature dependence. At temperatures up to 320 K majority of charge carriers are electrons, while at temperatures of 320 - 380 K protons start to play a significant role. \cite{4}

The coefficient of the quadratic nonlinearities of meta-nitroanilin can be found in \cite{3}. We find

\[\chi^{(2)} = 1.0005\times10^{-11} \mbox{ m/V} \]

For a rough estimate of the mean free path of the phonons, following the method proposed in \cite{6}:

\[l_{ph} \sim \frac{Mu^{2} d}{kT} \]

\noindent where $M$ - mass of the molecule, $u$ - the speed of sound, $d$ - the lattice period, $k$ -- Boltzmann constant,  $T$ -- the temperature.

Speed of sound in the organic molecular crystal is usually from 1000 to 10000 m/s . The numerical calculation of the molecular dynamics method gives $u \sim$ 6000 m/s. We take the lattice period $d = c =$ 5.082 \AA for the direction of the quadratic nonlinearity. With that said:

\[l_{ph} \sim 1\times10^{-6} \mbox{ m} \]

\noindent Permittivity at temperatures around 300 K according to \cite{4} is:

\[\varepsilon \approx 10 \]

Substituting the data available in the formula \eqref{2} we get:

\[ E^* \sim 156. \mbox{ V/m} \]

This means that the asymmetric component of the I-V curve of the sample is comparable with the current caused by such an external field.


\begin{thebibliography}{7}

\bibitem{1}
V.M.Gurevich, I.S.Zheludev, Conductivity anisotropy of a single crystal of lithium sulfate Li${}_{2}$SO${}_{4}$ $\cdot$ H${}_{2}$O // Crystallography Reports, 1960.

\bibitem{2}
I.F.Kanaev, V.K.Malinovsky, Asymmetry along the polarization axis conductivity in ferroelectric crystals // Proceedings of the Russian Academy of Sciences, 1982, vol. 266, num. 6.

\bibitem{3}
Edited by D.S.Chemla \& J.Zyss, Nonlinear Optical Properties of Organic Molecules and Crystals, Academic Press, Orlando, 1987.

\bibitem{4}
M.Szostak, H.Chojnacki, E.Staryga, M.Dluzniewski, G.Bak, Contribution to molecular mechanism of optical nonlinearity and electric conductivity of 3-nitroaniline single crystals by dielectric, electric and quantum chemical studies // Chemical Physics 365 (2009), p. 44-52.

\bibitem{5}
A.C.Skapski, J.L.Stevenson, X-Ray Crystal Structure of the Electro-optic Material meta-Nitroaniline // J. Chem. Soc., Perkin Trans. 2 (1973) 1197. 

\bibitem{6}
L.P.Pitaevskii, E.M.Lifshitz, Physical Kinetics (Course of Theoretical Physics, Volume 10), Pergamon Press, 1981.

\bibitem{7}
M.Pope, C.E.Swedberg, Electronic Processes in Organic Crystals, Clarendon, Oxford, 1982.

\end{thebibliography}
\end{document}

%%% remove comment delimiter ('%') and select language if required
\selectlanguage{russian} 

\title{О некоторых возможных особенностях движения полярона в гиротропной среде}

\author{Александр Северин \thanks{Автор благодарит Игоря Тодощенко за обсуждение и помощь в подготовке английского варианта статьи.} \\ \texttt{severin@kiam.ru}}

\begin{abstract}
В статье теоретически предсказывается возможность асимметричной динамики полярона в гиротропной среде и дается приближенная количественная оценка эффекта.
\end{abstract}

\maketitle

Существует целый класс явлений переноса заряда в твердом теле, который теоретически возможен, но до сих пор почти не изучен экспериментально. Это электропроводность, асимметричная относительно обращения направления. Блоховские электроны оказывают слабое обратное воздействие на кристаллическую решетку, поэтому их движение можно считать движением в потенциальном поле, и поэтому в случае зонной проводимости электропроводность в противоположных направлениях одинакова. Но это не относится к ионному и поляронному механизмам проводимости. Ионы вызывают значительную деформацию решетки из-за своей большой массы, а поляроны деформируют решетку по определению. Поэтому в тех случаях, когда заряд переносится ионами или поляронами, электропроводность в противоположных направлениях может быть различной.

За всю историю экспериментов удалось найти лишь два случая, когда гомогенная среда обнаруживала диодные свойства: \cite{1} и \cite{2}. Но в работе \cite{2} авторы объясняют эффект наличием сегнетоэлектрических микродоменов, то есть гомогенность на самом деле кажущаяся. В работе \cite{1} обнаружено, что электропроводность моногидрата сульфата лития вдоль одной из осей несимметрична относительно обращения направления. Отличие составляет около 10\%. Везде, кроме излома в начале координат, вольт-амперная характеристика линейна. Носителями заряда, вероятнее всего, являются ионы водорода, перемещающиеся по гротгусовскому (эстафетному) механизму. 

Но асимметричные динамические свойства поляронов пока еще нигде не были экспериментально зарегистрированы. Попробуем определить в каком веществе это явление может быть обнаружено. Прежде всего кристаллическая структура этого вещества должна быть гиротропной, то есть должно существовать хотя бы одно направление, такое, что будучи зеркально отраженной от плоскости, перпендикулярной этому направлению, никакими поворотами и перемещениями структура не может быть переведена сама в себя. Это те же самые требования к симметрии, что и у пироэлектриков, и им соответствуют следующие 10 точечных групп, не обладающих центром симметрии:  1, 2, 3, 4, 6, m, mm2, 3m, 4mm, 6mm. Скорее всего, это будет молекулярный кристалл, поскольку в таких кристаллах перенос заряда часто осуществляется поляронами малого радиуса.

Асимметрия взаимодействия полярона со средой может проявляться по-разному, но проще всего количественно оценить асимметрию, вызванную квадратичной нелинейностью диэлектрической поляризации.

В нелинейной диэлектрической среде поляризация $p$ определяется формулой:

\[p=\chi E+\chi ^{(2)} E^{2} +\chi ^{(3)} E^{3} +...\]

\noindent где $E$ - напряженность электрического поля, $\chi$, $\chi^{(2)}$, $\chi^{(3)}$ - линейная, квадратичная и кубическая поляризуемости. Материалы с квадратичной и кубической нелинейностями поляризации широко используются в оптике для удвоения и суммирования частот лазерных лучей. Квадратичной нелинейностью обладают все пироэлектрики.

Легко заметить, что если заряд находится в линейной диэлектрической среде, то силы взаимодействия заряда с диполями среды полностью уравновешиваются. Но если среда квадратично-нелинейна - то нет.

Сила взаимодействия полярона с квадратично-нелинейной средой равна:

\[F=\int _{R}\chi ^{(2)} E_{x}^{2} \frac{\partial E_{x} }{\partial x} dr \]

\noindent где $R$ - область пространства снаружи полярона, $x$ - направление, в котором среда квадратично-нелинейна.

Если считать, что линейная составляющая поляризуемости изотропна, и полярон имеет сферическую форму, то:

\begin{equation} \label{1} 
F=\frac{4\pi }{15} \left(\frac{e}{4\pi \varepsilon \varepsilon _{0} } \right)^{3} \frac{\chi ^{(2)} }{r_{0}^{4} }  
\end{equation}

\noindent где $e$ - элементарный заряд, $\epsilon$ - диэлектрическая проницаемость среды, $\epsilon_{0}$ - диэлектрическая постоянная, $r_{0}$ - радиус полярона.

Поскольку действие силы \eqref{1} на единичный полярон аналогично действию внешнего электростатического поля, целесообразно ввести величину $E^*=F/e$ и назвать ее квазинапряженностью. Тогда:

\begin{equation} \label{2} 
E^{*} =\frac{4\pi }{15} \frac{e^{2} }{\left(4\pi \varepsilon \varepsilon _{0} \right)^{3} } \frac{\chi ^{(2)} }{r_{0}^{4} }  
\end{equation}

Разумеется, формула \eqref{1} может быть точной только для поляронов большого радиуса. Первая и самая очевидная причина состоит в том, что на расстояниях порядка периода решетки мы не можем рассматривать диэлектрик как однородную среду.

Но есть и другая, не менее важная. Перемещение полярона в соседнюю локацию происходит за счет тепловых флуктуаций решетки, в результате которых исчезает или уменьшается потенциальный барьер между локациями. Такая флуктуация должна сильно влиять на распределение зарядов вблизи полярона. Не уничтожит ли она асимметрию, возникшую в результате поляризации среды?

Пусть энергия полярона в данной локации в отсутствии колебаний решетки - $W_{1}$, а энергии в двух соседних локациях, находящихся в противоположных направлениях - $W_{0}$ и $W_{2}$. В кристалле, обладающем центром симметрии $W_{0}=W_{2}$. Но в нецентросимметричном кристалле из-за описанного выше эффекта $W_{0}\neq W_{2}$, точнее $W_{0}-W_{2}= 2Fa$, где $a$ - период решетки.

Сейчас нас интересует сравнительная вероятность флуктуаций, перебрасывающих полярон из локации 1 в локации 0 и 2. Если бы эта разность энергий была создана внешним полем, мы сказали бы, что перескок в локацию 2 более вероятен, поскольку энергетический барьер меньше. 

Но здесь разность энергий создается зарядами самой решетки, которые участвуют в колебаниях. Можно предположить, что асимметрия поляризации вблизи полярона исчезнет, но поле, создаваемое частью решетки, лежащей на каком-то достаточно большом расстоянии, будет действовать как внешнее.

Равновесные состояния полярона в локациях 0, 1 и 2 имеют одинаковую энергию в силу трансляционной инвариантности. Переход полярона из локации 1 в локацию 0 представляет собой обращенный по времени переход из локации 1 в локацию 2. Если пренебречь рассеянием фононов, то поле межатомного взаимодействия можно считать потенциальным. Также можно считать потенциальным действующее на атомы поле электрона, в результате локализации которого образовался полярон, поскольку масса электрона много меньше массы атома, и поэтому в рамках данного приближения состояние электрона можно считать функцией координат атомов. Таким образом, в рамках принятого приближения прыжок полярона представляет собой движение в потенциальном поле, которое, как известно, обратимо. Следовательно, прыжки в локации 0 и 2 потребуют одинаковой энергии и будут одинаково вероятны. 

Асимметрия возможна лишь за пределами описанного приближения, то есть если не пренебрегать рассеянием фононов и массой электрона. Первый из этих факторов удобно учитывать при помощи длины свободного пробега фонона $l_{ph}$.

Фононы, возникшие в результате попадания полярона в новую локацию, не выходят за пределы сферы радиусом $l_{ph}$. Фононы, выбивающие полярон из старой локации, должны родиться в результате флуктуации внутри сферы радиусом $l_{ph}$. Таким образом, весь процесс прыжка полярона, который мы выше приближенно считали обратимым, локализован внутри сферы радиусом $l_{ph}$, а среду за ее пределами можно рассматривать как источник внешнего поля. Поэтому мы можем использовать формулы \eqref{1} и \eqref{2} в качестве грубой оценки для поляронов малого радиуса, если вместо радиуса полярона $r_{0}$ подставить длину свободного пробега фонона $l_{ph}$.

Оценим величину эффекта в конкретном веществе. Выберем для этого метанитроанилин (3-нитроанилин, $\mu$-нитроанилин) - C${}_{6}$H${}_{6}$N${}_{2}$O${}_{2}$. Метанитроанилин кристаллизуется в виде молекулярного кристалла, по четыре молекулы в ячейке, решетка орторомбическая, размеры элементарной ячейки $a$ = 6.501 \AA, $b$ = 19.33 \AA, $c$ = 5.082 \AA. \cite{5}

Монокристалл метанитроанилина обладает высокой квадратичной нелинейностью. Механизм электропроводности прыжковый, что доказывается её температурной зависимостью. При температурах до 320 K основными носителями заряда являются электроны, при температурах 320 - 380 K начинают играть существенную роль протоны. \cite{4}

Коэффициент квадратичной нелинейности метанитроанилина можно найти в \cite{3}, с. 268. Отсюда:

\[\chi^{(2)} = 1.0005\times10^{-11} \mbox{ м/В}\]

Для грубой оценки длины свободного пробега фононов, используем способ, предложенный в \cite{6}, с. 352.

\[l_{ph} \sim \frac{Mu^{2} d}{kT} \]

\noindent где $M$ - масса молекулы, $u$ - скорость звука, $d$ - период решетки, $k$ -- постоянная Больцмана, $T$ - температура.

Скорость звука в молекулярных органических кристаллах обычно составляет от 1000 до 10000 м/с. Численный расчет методом молекулярной динамики дает $u \approx$ 6000 м/c. Период решетки возьмем $d = c =$ 5.082 \AA, поскольку именно в этом направлении наблюдается квадратичная нелинейность. С учетом сказанного:

\[l_{ph} \sim 1\times10^{-6} \mbox{ м} \]

Диэлектрическая проницаемость при температурах около 300 К согласно \cite{4}:

\[\varepsilon \approx 10.\] 

Подставив имеющиеся данные в формулу \eqref{2} получим:
 
\[E^{*} \sim 156. \mbox{ В/м}\]

Это означает, что асимметричная составляющая вольт-амперной характеристики образца сравнима с током, вызываемым таким внешним полем.